\documentclass[conference]{IEEEtran}
\IEEEoverridecommandlockouts

\usepackage{cite}
\usepackage{amsmath,amssymb,amsfonts}
\usepackage{algorithmic}
\usepackage{graphicx}
\usepackage{textcomp}
\usepackage{xcolor}
\def\BibTeX{{\rm B\kern-.05em{\sc i\kern-.025em b}\kern-.08em
    T\kern-.1667em\lower.7ex\hbox{E}\kern-.125emX}}

\date{}

\usepackage{tikz}
\usepackage{amsmath}

\usepackage[normalem]{ulem}

\usepackage{cite}
\usepackage{amsmath,amssymb,amsfonts}
\usepackage{algorithmic}
\usepackage{graphicx}
\usepackage{textcomp}
\usepackage{xcolor}
\def\BibTeX{{\rm B\kern-.05em{\sc i\kern-.025em b}\kern-.08em
    T\kern-.1667em\lower.7ex\hbox{E}\kern-.125emX}}

\usepackage{subfiles}
\usepackage{subfig}
\usepackage{microtype}
\usepackage{verbatim}
\usepackage[utf8]{inputenc}
\usepackage[english]{babel}
\usepackage{amsthm}
\usepackage{colortbl}
\usepackage{multirow}
\usepackage{tablefootnote}

\usepackage{tikz}
\usepackage{enumitem}
\usepackage{fancyhdr}

\usepackage{xurl} 
\usepackage{multicol}

\newcommand*\circled[1]{\tikz[baseline=(char.base)]{
            \node[shape=circle,draw,inner sep=2pt] (char) {#1};}}

\newcommand{\fulln}{Zipline}
\newcommand{\shortn}{Zipline}
\newcommand{\rpprod}{QP\textsubscript{prod}}
\newcommand{\rpcon}{QP\textsubscript{con}}

\newcommand{\tms}{$\times$}
\newcommand{\myparagraph}[1]{\vspace{1em}\noindent {\bf #1:}}

   \newcommand*{\RELEASE}{}
\definecolor{brass}{rgb}{0.71, 0.65, 0.26}
\ifdefined\RELEASE
    \newcommand\old[1]{}
    \newcommand\boris[1]{}
    \newcommand\dmitrii[1]{#1}
    
    \newcommand\marios[1]{}
    \newcommand\shyam[1]{}
    \newcommand\artemiy[1]{}
    \newcommand\update[1]{}

\else
    \newcommand\old[1]{}
    \newcommand\boris[1]{{\color{cyan}[B]: #1}}
    \newcommand\dmitrii[1]{{\color{blue}[D]: #1}}
    
    \newcommand\marios[1]{{\color{cyan}[M]: #1}}
    \newcommand\shyam[1]{{\color{green}[S]: #1}}

    \newcommand\update[1]{{\color{red}[UPDATE]: #1}}

\fi

\begin{document}

\title{
Shattering the Ephemeral Storage Cost Barrier \\ for Data-Intensive Serverless Workflows
}





\author{
\\{\rm Dmitrii Ustiugov\footnotemark*\thanks{* The work was done when the author was at the University of Edinburgh}}\\
NTU Singapore
\and
\\{\rm Shyam Jesalpura}\\
University of Edinburgh
\and
\\{\rm Mert Bora Alper\footnotemark*{}}\\
Stripe
\and
\\{\rm Michal Baczun\footnotemark*{}}\\
Bloomberg
\and
\\{\rm Rustem Feyzkhanov}\\
Instrumental
\and
\\{\hspace{60pt}\rm Edouard Bugnion}\\
\hspace{60pt}EPFL
\and
\\{\rm Marios Kogias}\\
Imperial College London \&\\
Microsoft Research
\and
\\{\rm Boris Grot}\\
University of Edinburgh 
} 
\vspace{10pt}



\date{}
\maketitle


\begin{abstract}

Serverless computing has emerged as a popular cloud deployment paradigm. In serverless, the developers implement their application as a set of functions forming a workflow in which functions invoke each other. The cloud providers are responsible for automatically scaling the number of instances for each function on demand and forwarding the requests in a workflow to the appropriate function instance.
Problematically, today's serverless clouds lack efficient support for cross-function data transfers in a workflow, preventing the efficient execution of data-intensive serverless applications. 
Instead, functions transmit intermediate, i.e., ephemeral, data to other functions through third-party services, such as AWS S3 storage, AWS ElastiCache in-memory cache, or multi-tier solutions proposed by prior works. 

We show that data-intensive application deployments in serverless platforms that rely on such {\em through-storage} data transfers are economically impractical, with the storage costs accounting for $>$24-99\% of the total serverless execution bill.
We introduce \fulln, an API-preserving fast data communication method for serverless that enables direct function-to-function transfers. 
With \shortn{}, the sender function's runtime buffers the payload in its memory and sends a reference to the receiver, which the load balancer and autoscaler pick based on the current load. Using the reference, the receiver instance pulls the transmitted data directly from the sender's memory. \shortn{} is natively compatible with existing autoscaling infrastructure, preserves function invocation semantics, and avoids the cost and performance overheads of using an intermediate service for data transfers.
We prototype our system in vHive/Knative deployed on a cluster of AWS EC2 nodes, showing that \shortn{} reduces the overall cost {\em and} improves latency and bandwidth over AWS S3 (lowest cost state-of-the-art) and ElasticCache (highest performance state-of-the-art). 
On real-world applications, \shortn{} reduces the overall cost of 2-5$\times$ {\em and} application execution time by 1.3-3.4$\times$ compared to S3. Compared to ElastiCache, \shortn{} reduces the total cost by 17-772\tms{} while improving performance by 2-5\%.
\end{abstract}

\section{Introduction}
\label{sec:intro}

Serverless computing has emerged as a pervasive cloud technology due to its scalability, resource- and cost-efficiency~-- factors that benefit both cloud providers and their customers.
All major cloud providers have serverless offerings, including AWS Lambda~\cite{agache:firecracker}, Azure Functions~\cite{azure:functions}, and Google Cloud Functions~\cite{google:functions} and Google Cloud Run~\cite{google:cloudrun}.
In serverless, the application logic is organised as a set of {\em stateless functions} that communicate with each other and with cloud storage services hosting the application state.
Serverless computing is expressive enough to support many applications, e.g., video encoding~\cite{romero:llama,fouladi:encoding}, compilation~\cite{klimovic:understanding,fouladi:laptop} and machine learning~\cite{jiang:demystifying}.


The stateless and ephemeral nature of function instances mandates that functions communicate any intermediate and ephemeral state across the functions comprising the application logic. Inter-function communication generally happens when one function, the {\em producer}, invokes one or more {\em consumer} functions in the workflow and passes inputs to them. Crucially, the instances of the consumer functions are not known by the producer at invocation time because they are picked by the cloud provider's load balancer and autoscaler components on demand. Also, for many applications, the amount of data transmitted across function instances can be large, measuring 10s of MBs or more; examples include video analytics~\cite{fouladi:encoding,fouladi:laptop,romero:llama,romero:faast}, data analytics~\cite{pu:shuffling,muller:lambada,perron:starling}, and ML~\cite{jiang:demystifying}.


Programming model for data communication is {\em object-centric}, with the functions passing data via an intermediate external service, typically using a \texttt{put()/get()} interface. This intermediate service
can be a storage service (e.g., AWS S3 or Google Cloud Storage) or an in-memory cache service (e.g., AWS ElastiCache), which requires the producer function to first store the data, then invoke the consumer, and subsequently have the consumer retrieve the data from storage. 
The indirection via an intermediate service introduces large latency overheads and adds to the cost of the intermediate service. We further refer to such services as {\em storage services}.


Researchers have identified the problem of efficient serverless communication and have proposed solutions. Some seek to improve the performance of storage-based transfers using tiered storage, such as combining an in-memory cache layer (e.g., ElastiCache)  with a cold storage layer (e.g., S3)~\cite{mahgoub:sonic,sreekanti:cloudburst,romero:faast,mvondo:ofc}. While tiered storage can improve performance over a single storage layer (or cost over a single in-memory cache layer), the disadvantages of through-storage indirection remain. 

We observe that serverless architectures that use through-storage transfers, whether via AWS S3 or previously proposed intelligent multi-tier services, are economically unattractive due to the prohibitively high costs of storing transmitted data in storage services. 
We find that even if these services implemented perfect garbage collection, i.e., deallocating transmitted objects immediately after the last retrieval, the costs of intermediate 
bookkeeping would still dominate the overall execution cost for data-intensive applications. For example, we show that transmitting data via S3 and ElastiCache in a MapReduce application's shuffle phase can account for 70\% to over 99\% of the total processing cost.

By studying the production traces from Azure Functions~\cite{romero:faast}, we make the following key observation: 75\% of data objects transmitted across functions must be buffered for only 30 seconds or less. In contrast, a function instance's lifetime (e.g., the minimum keep-alive period of an idle instance before serverless infrastructure tears it down) spans to many minutes~\cite{agache:firecracker,shahrad:serverless}. Hence, the function instances' lifetime significantly exceeds the transmitted data's lifetime.

We exploit the disparity between the data and instances' lifetimes and introduce \fulln{}\footnote{We plan to release the \shortn's source code by the time of publication.}, a serverless communication substrate that allows direct communication between two function instances in a manner that is flexible and compatible with the autoscaling infrastructure used by cloud providers. \shortn{} preserves the existing API and invocation semantics of serverless functions while avoiding the need for intermediate storage for arbitrarily-sized data transfers. At the heart of \shortn{} is an explicit separation of the control plane used for function invocation, which is tightly integrated with the autoscaling infrastructure, from the data transfer itself. In simplest terms, with \shortn{}, the producer function buffers the data that needs to be transferred in its memory and sends a reference to the data inlined with the invocation to the consumer function. The consumer then directly {\em pulls} the data from the producer's memory. 
More concretely, \shortn{} defines a short-lived namespace of objects with the same lifetime as the function instance. Subsequent function instances can access this namespace through references that do not expose the underlying infrastructure to the user code. 


\shortn{} naturally supports a variety of inter-function communication patterns, including producer-consumer, scatter (map), gather (reduce), and broadcast.
Compared to through-storage transfers, \shortn{} avoids high-latency data copies to and from a storage layer and the associated monetary cost of storage usage. Critically, \shortn{} is fully compatible with the autoscaling infrastructure and requires minimal modifications at the endpoints of the existing control plane. 

We prototype \shortn{} in Knative~\cite{knative}, which is used in many commercial clouds~\cite{knative-offerings}, by extending its queue-proxy components with \shortn{} support.
We evaluate our proposal by deploying a \shortn-enabled vHive cluster in AWS EC2.
Using real-world applications, we show that \shortn{} delivers 2-5\tms{} lower cost {\em and} superior performance versus
transfers via S3 storage (i.e., cheapest among existing solutions) and ElastiCache in-memory cache (i.e., fastest among prior works) 
for all the above communication patterns in serverless computing.


The main contributions of our work are as follows:

\begin{itemize}[leftmargin=0cm,itemindent=.4cm,labelwidth=\itemindent,labelsep=0cm,align=left]
\setlength\itemsep{0pt}
    \item We show that through-storage inter-function communication imposes prohibitively high storage-related costs, accounting for 24-99\% of the total expenses, dominating the overall serverless execution bill for data-intensive applications.

    \item We find that the lifetime of serverless function instances significantly exceeds the lifetime of the data objects transmitted across functions, indicating the opportunity for using the instances' memory for buffering the transmissions.
    
    \item We introduce \shortn{}, which uses control/data path separation to pass an object reference to a consumer function instance as part of an invocation request, and delegates to the consumer, pulling the data from the producer's memory. \shortn{} supports various inter-function communication patterns and is fully compatible with serverless autoscaling infrastructure.  

    \item We demonstrate that \shortn{} is flexible and fast. On real-world applications, \shortn{} outperforms S3 by 1.3-3.4\tms{} with cost savings of 2-5\tms{}. 
    Furthermore, \shortn{} also consistently outperforms ElastiCache (the highest-performance data transfer option available today) by 2-5\% while slashing the overall application execution cost by 17-772\tms{}.

    

\end{itemize}

\section{Background and Motivation}
\label{sec:back}

Below we describe the modern serverless cloud architectures and programming models for data-intensive applications and evaluate the associated performance and cost overheads. 

\subsection{Serverless Computing and Autoscaling}
\label{sec:serverless_101}

The division of labor between the programmer and the underlying infrastructure is central to the serverless paradigm, defining its programming model and infrastructure management philosophy. The programming model defines {\em a function} as the key abstraction for programming and {\em a function instance} as a unit of placement and scaling from the infrastructure perspective. Using these abstractions, developers can write and deploy arbitrary applications without considering system configuration and cloud resource management. Instead, serverless infrastructure transparently manages cloud resources, continuously adjusting the number of function instances based on the current function invocation traffic. 

\begin{figure}[t!]
    \centering
    \includegraphics[width=\columnwidth]{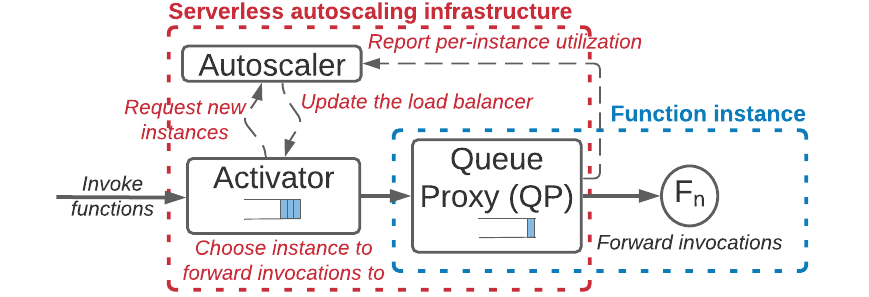}
    \caption{Operation of serverless autoscaling infrastructure.
    }
    \vspace{-15pt}
    \label{fig:autoscaling_infra}
\end{figure}

We describe the operation of serverless autoscaling infrastructure~(Figure~\ref{fig:autoscaling_infra}) using the Knative~\cite{knative} terminology since it is used widely in production~\cite{knative-offerings} and representative of the leading clouds\cite{liu:gap}.
The autoscaling infrastructure of serverless aims to achieve two objectives. The first objective is responding to load changes by spawning new function instances when the load increases and shutting down idle instances once the load drops. The second objective is minimizing queuing latency by balancing the load across the active instances.

Instance scaling and load-balancing decisions inherently rely on utilization metrics from the active function instances, gathered and stored with the help of the following two components. 
Each function invocation traverses a provider-managed {\em queue-proxy} component, which is in charge of forwarding incoming requests to the function instance with which the queue proxy is co-located.
Queue proxy also collects and reports utilization metrics of that instance to the {\em autoscaler} control plane component.
The {\em autoscaler} monitors the load in front of active instances and implements the scaling policy.

To balance the load among all active instances of a function, serverless clouds employ a {\em load balancer} whose job is to steer each request to one of the instances. Every request must traverse the load balancer, referred to as the {\em activator} in Knative. 
The autoscaler periodically updates the activator about active instances and their load.
If there is an incoming request for a function and no active instances are available, or all of them are busy, the activator needs to request new instances of that function from the autoscaler.
The autoscaler makes a placement decision and spawns a new instance while the activator buffers the pending request.
Once the instance is up, the activator steers the invocation to the instance via its corresponding queue proxy. 

Together, queue proxy, autoscaler, and load balancer drive serverless functions autoscaling. The rest of the system is designed around this triplet to deliver scalability to application developers and resource efficiency to cloud providers.

\subsection{Programming Data-Intensive Applications in Serverless }
\label{sec:data-intens}

Data-intensive applications are ubiquitous in today's serverless clouds~\cite{fouladi:encoding,fouladi:laptop,romero:llama,romero:faast,pu:shuffling,muller:lambada,perron:starling,jiang:demystifying}. However, these applications are challenging to support with contemporary serverless infrastructure because
they need to rapidly communicate state from one processing stage (i.e., function) to another in an application workflow.
Typically, serverless functions represent a single stage (e.g., map and reduce functions) in such an application's data-processing workflow while each function's instances can be in charge of a single piece of state~\cite{fouladi:laptop,klimovic:pocket,klimovic:understanding}.
This way the developer is completely free of any autoscaling or resource management decision since the application can leverage any available compute resource offered to it depending on the combination of pieces of state and workflow stages.

The challenge is to devise a solution for fast communication of these pieces of state, further referred to as \textit{objects}, across different functions (i.e., their instances) while keeping the benefits of serverless computing, i.e. elasticity, and without substantially increasing cost.
Establishing direct communication between the instances using traditional POSIX APIs, e.g., via sockets, could enable cheap and high-performance communication. Still, it would require the developer to devise custom autoscaling and data-partitioning solutions in the application code. Thus, such solutions forego serverless computing's main advantage, which is its obviating the need for infrastructure management by application developers.
Instead, existing data-intensive applications on serverless pass objects across functions via storage services, such as cloud storage or in-memory cache, typically using a \texttt{put()/get()} API~\cite{romero:faast,eismann:state}.\footnote{In serverless, it is possible to pass small data objects inline in simple workflows such as a pipeline.
For example, AWS Lambda supports inline transfers for objects smaller than 256KB and 6MB for asynchronous and synchronous function invocations, respectively. Thus, through-storage transfers are dominant in practice.}
We refer to these communication methods as {\em through-storage} transfers. 

\subsection{Through-Storage Transfers and Their Cost}
\label{sec:storage-and-cost}



Through-storage communication imposes a performance overhead as well as a financial cost. The financial aspect includes the cost for each \texttt{Get()} or \texttt{Put()} operation plus the cost of storing data objects in remote storage in GB-hour. We refer to the latter as the {\em storage lease} cost, proportional to the time and space the objects take up in the remote storage. 

Prior works have rigorously explored the cost-performance trade-offs in storage architectures for data-intensive serverless applications using a conventional storage service as the cheapest option, an in-memory cache service as the highest-performance alternative, or a multi-tier combination of the two. For example, Locus~\cite{pu:shuffling} uses
different storage tiers for specific purposes, namely AWS ElastiCache for shuffling and
S3 for cold storage. Pocket~\cite{klimovic:pocket} and SONIC~\cite{mahgoub:sonic} 
employ a similar idea and develop a control-plane solution to multiplex different
storage services based on inferred application needs. Faa\$T\cite{romero:faast}, 
Cloudburst~\cite{sreekanti:cloudburst}, and OFC~\cite{mvondo:ofc} propose using
key-value stores to cache objects in the main memory or disk distributed across the serverless cluster deployment. Since all these solutions require additional bookkeeping, their usage adds to the bill of serverless application execution in two ways. The first cost type is associated directly with the storage usage, namely the cost of operations and storage lease. The second cost type is indirect, as the latency of saving and retrieving objects from storage is a part of the billed serverless function execution time, i.e., the computational cost. Hence, transferring data through slow storage would result in higher computational costs.

\begin{figure}[t]
    \centering
    \includegraphics[width=\columnwidth]{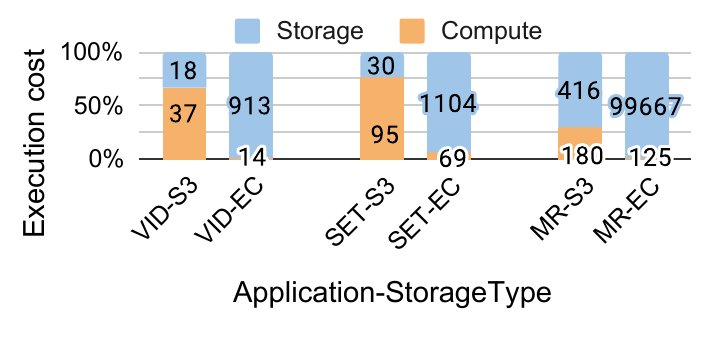}
    \vspace{-25pt}
    \caption{The cost breakdown for real-world data-intensive multi-function applications~(\S\ref{sec:method_apps}), namely Video Analytics~(VID), Stacking Ensemble Training~(SET), and Map-Reduce~(MR), when performing data transfers through AWS S3 and ElastiCache~(EC). The numbers show the cost values in (in $USD\times10^{-6}$) for compute and storage expenses.
    }
    \vspace{-10pt}
    \label{fig:back-cost}
\end{figure}

Surprisingly, even using the cheapest storage technologies may dominate the overall execution cost for data-intensive serverless applications. We estimate the costs from an application developer's perspective based on the applications' execution time measured in a real-world deployment and using the AWS pricing model to determine the billed execution time and storage capacity~\cite{aws:pricing,s3:pricing,elasticache:pricing} (details are in
\S\ref{sec:method}).
In AWS Lambda, compute time is billed proportionally to execution time and maximum function memory usage. We estimate the lower-bound costs for buffering transferred data objects in AWS S3 and ElastiCache, making two conservative assumptions:  (1) data objects are de-allocated immediately after the last retrieval and (2) storage space is never overprovisioned.\footnote{In practice, services rarely satisfy these assumptions. For example, as of 2024, AWS S3 lacks support for expiration time below one day. Newly added AWS Serverless ElastiCache is metered for a minimum of 1 GB of data stored~\cite{elasticache:pricing}. Hence, storage costs would account for a larger fraction of expenses in real-world deployments.}
Figure~\ref{fig:back-cost} shows the cost breakdown for three widely-used applications, accounting for 24-70\% and 94-99\% of the overall cost when using S3 and ElastiCache for transfers, respectively.

Given the above results, it is clear that serverless architectures that rely on through-storage transfers are economically infeasible for data-intensive applications.  
Although multi-tier proposals~\cite{mahgoub:sonic,klimovic:pocket,mvondo:ofc,romero:faast} have narrowed the performance gap between in-memory caches and cloud storage, the latter still underpins their most cost-optimal tier, hence such systems would impose much storage costs in between our estimates for S3 and ElastiCache based transfers. Finally, the slowest tier in multi-tier systems inevitably becomes the main source of the tail latency increase for data-intensive applications, as discovered by prior work~\cite{ustiugov:analyzing}.

\section{\fulln{} Communication}
\label{sec:design}

\subsection{Design Insights}
\label{sec:design-insights}

We exploit three insights that enable a serverless communication model, which, in the common case, obviates the need for through-storage transfers.
Our first insight is to separate control (function invocation) and data (transfer) paths without impacting the functioning of the autoscaling infrastructure.
The challenge is doing so without resorting to a storage service, which is what existing through-storage transfers rely on. We address this challenge with the help of the second insight.

\begin{figure}[t]
        \includegraphics[width=0.85\columnwidth]{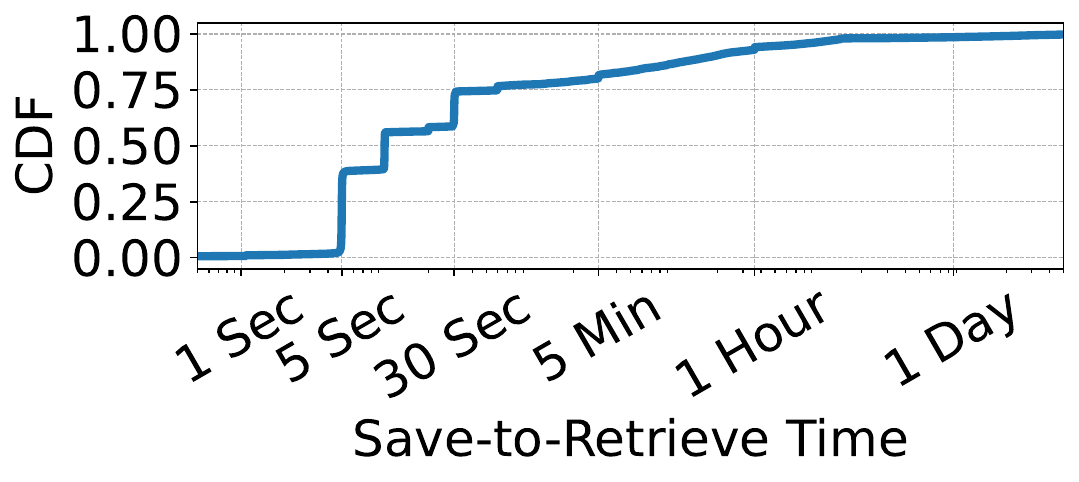}
        \label{fig:nonweighted_blob_cdf}
\centering
\vspace{-5pt}
\caption{Cumulative Distribution Functions (CDFs) of the time duration between saving a data object in storage and its last retrieval, based on Azure Blob Traces~\cite{romero:faast}. Note the logarithmic scale on the horizontal axis.
}
\vspace{-10pt}
\label{fig:blob_cdf}
\end{figure}

The second insight is that the data transferred between instances are ephemeral, with lifetimes on the order of a few seconds.
Using the Azure Blob Traces~\cite{romero:faast}, we analyze the time between an object produced by one function and its {\em last} retrieval by another function of the same application. Figure~\ref{fig:blob_cdf} shows that 75\% of the data objects transferred across functions are consumed within 30 seconds.
Hence, the data lifetime is much shorter than the keep-alive period of serverless functions (which is typically in the order of minutes to maximize the likelihood of a warm invocation~\cite{agache:firecracker, shahrad:serverless}). 


Based on the above, we draw one final insight: instead of using a storage service to communicate data across function instances, a producer instance can simply buffer the data in its own memory and have the consumer instance pull from it. 
We note that most language runtimes require buffering the transmitted object in a memory buffer before calling the \texttt{Put()} API of the storage service, so the system only needs to provide a way to pass a pointer to that buffer to the target consumer instance.
This insight forms the foundation for \shortn{}, presented next.




\subsection{Design Overview}
\label{sec:design-overview}


We introduce \fulln{}, a serverless-native data communication fabric that meets all five serverless communication requirements: high performance, compliance with the existing semantics of serverless function invocations, compatibility with autoscaling, and the standard data-transfer API in serverless. 

Following the insights developed in Sec.~\ref{sec:design-insights}, {\shortn} splits the function invocation plane into control and data planes. Crucially, the control plane is unchanged, matching the existing serverless architecture (Figure~\ref{fig:autoscaling_infra}), thus allowing the autoscaling infrastructure to take the load balancing decisions for each incoming invocation by steering the invocation to the least-loaded instances of a function. The control plane carries only the function invocation control messages, i.e., RPCs. The data plane is responsible for transferring the objects. 

In simplest terms, a producer function instance in \shortn{} buffers the data to be communicated to the consumer(s) in its own memory and sends a reference to the data inline with the invocation to the consumer function(s). The consumer(s) then directly {\em pull} the data from the producer's memory. 
\shortn{} fundamentally replaces push-based data transfers, in which the producer pushes the data through the activator or through a storage layer, with an approach in which the consumer directly pulls the data after the control plane has made its decisions.

\begin{figure}[t!]
    \centering
    \includegraphics[width=\columnwidth]{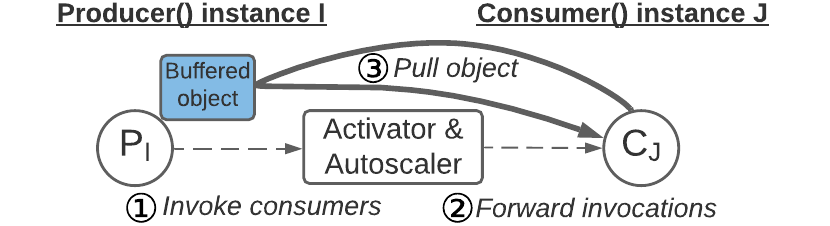}
    \caption{\shortn{} architecture overview. 
    }
    \label{fig:xdt-design}
\end{figure}

Figure~\ref{fig:xdt-design} describes {\shortn} operation. Let us assume two serverless functions, a producer and a consumer, each of which may have any number of instances at any point in time. As in the case with existing communication methods, the producer logic invokes the consumer function while passing a data object as an argument. However, in contrast to the existing systems, in \shortn{}, consumer function invocations travel to the activator separately from their corresponding objects~\circled{1}, which remain buffered at their source. After contacting the autoscaler as needed, the activator chooses the instance of the consumer function, to which the activator forwards the invocation for processing~\circled{2}. Once the invocation arrives at the target instance, the instance can pull the object from the producer instance~\circled{3}, using the reference enclosed in the invocation message.





\begin{table}
\centering
\begin{tabular}{ll}
\hline
\multicolumn{1}{l}{API Call} & \multicolumn{1}{l}{Description}                   \\ 
\hline
\texttt{rsp := invoke(URL, obj)}    & Invoke a function          \\
\texttt{ref := put(obj, N)}             & Buffer an object locally  \\
\texttt{obj := get(ref)}                & Fetch a remote object \\
\hline
\end{tabular}
\caption{{\shortn} API description.}
\label{tbl:api}
\end{table}

\subsubsection{\shortn{} Programming Model}
\label{sec:xdt-api}

The \shortn{} programming model features a minimalist yet expressive API~(Table~\ref{tbl:api}) that supports all three essential communication patterns, namely invoking a function,
scattering and broadcasting objects to several consumers, and gathering the output of several functions. The \shortn{} API is fully compatible with the API supported by production clouds, such as AWS Lambda and S3's Boto3~\cite{boto3}.

First, \shortn{} supports the standard blocking API, as in AWS Lambda~\cite{aws:invoke}, which is the \texttt{invoke()} call that invokes a function by its \texttt{URL}, passing a binary data object \texttt{obj} by value.
Upon invocation, the API of the \shortn{} SDK is responsible for buffering the object at the producer side until the consumer function instance, chosen by the autoscaling infrastructure, pulls it.
In this case, the consumer function starts processing {\em after} the object is transferred to the consumer instance.

\shortn{} also supports the standard non-blocking (asynchronous) interface, which is similar to a common key-value store interface like in AWS S3~\cite{boto3}, namely \texttt{get()} and \texttt{put()} calls.
In contrast to using a storage service, with \shortn{}, the sender instance of the producer function can finish the invocation before one of the consumer instances retrieves the transmitted object.

To de-couple the function invocation and the data transfer interfaces, \shortn{} introduces {\em \shortn{} references} as a first-class primitive. 
When the producer function calls \texttt{put()}, the runtime returns a \shortn{} reference to a specific object while retaining an immutable copy of the object.\footnote{Note that during non-blocking transfers, the producer function's user code allocates the object, with the \shortn{} SDK only holding references to it.}
When the consumer needs to read this object, it calls \texttt{get()} that pulls the object from the remote server.
Each reference is associated with a user-specified number of retrievals \texttt{N} of that object, which complete before the producer instance's runtime de-allocates the object. From the user perspective, references are just opaque hashes that do not expose any information regarding the underlying provider infrastructure, and that can be neither generated nor manipulated by user code.

A \shortn{} reference also includes a \shortn{} ID that unambiguously identifies objects created in the same workflow, even when an application's functions process many invocations concurrently.
For example, when multiple instances of the same function in a multi-stage (multi-function) video processing pipeline produce objects, the objects will be guaranteed to have different \shortn{} IDs. This mechanism is similar to the tracing implementation in distributed tracing frameworks~\cite{otel:traces}.

This programming model allows the seamless porting of serverless applications, e.g., those implemented for AWS Lambda or Knative serverless platforms, with corresponding wrapper functions. To demonstrate the API's portability, we implemented \shortn{} SDKs for applications written in Python and Golang, and deployed them in a Knative cluster.

\subsubsection{
\shortn{} Semantics \& Error Handling
}
\label{sec:lifecycle}


Function invocations in modern serverless offerings, like AWS Lambda and Azure Functions, provide the \textit{at-most-once} semantics~\cite{fox:harvest, kogias:r2p2, lee:rifl}, i.e., an invocation may execute not more than once even in the presence of a failure.\footnote{The user can construct primitives with at-least-once semantics by combining primitives with at-most-once semantics and re-try logic. Prior work also shows constructing primitives with the exactly-once semantics~\cite{lee:rifl}.}
Hence, the provider is responsible for exposing the runtime errors to the user logic to handle them~\cite{aws:step-errors,aws:errors,shilkov:durable,burckhardt:netherite}. Error handling logic in today's applications varies based on the function composition method.
The user can compose the functions as a direct chain (e.g., the producer makes a blocking call to the consumer) or chain the functions in an asynchronous workflow. In the latter case, an \textit{orchestrator} invokes the functions within the workflow. The orchestrator can be provider-based (e.g., AWS Step Functions~\cite{aws:step} and Azure Durable Functions~\cite{azure:durable}) or an auxiliary function that drives other functions. Handling failures may require 
re-execution of several functions. In this case, the first function of the sub-workflow must be re-invoked with the same arguments as the original invocation. Hence, the user is responsible for passing the first function's context throughout the sub-workflow to the function that can detect its failure.

Handling of \shortn{}-related failures follows the same approach. We describe a \shortn{} failure scenario in a two-function workflow with one producer function and one consumer function, which can be generalized to an arbitrary workflow. Crucially, the lifetime of a \shortn{} object is connected to the lifetime of the producer instance, thus a shutdown of a producer instance leads to immediate de-allocation of all the objects, retrievals of which have not completed.

For blocking invocations, i.e., the ones invoked with the \texttt{invoke()} call, the producer instance stays alive waiting for the response from the consumer, and may decide to re-invoke the consumer invocation if the previous invocation returns an error. For the non-blocking invocations, a \shortn{} transfer may fail if the producer instance is killed (e.g., due to exceeding the maximum invocation processing time) before a consumer instance retrieves the transmitted object. For example, the producer function may return success before the transfer is complete, followed by the instance shutdown. However, in this case, the consumer function receives the corresponding error when executing \shortn{} \texttt{get()}. 
The invocation of the consumer can follow the at-least-once semantics approach. To guarantee the correct execution of the entire workflow, the consumer must re-invoke the workflow starting from the producer function. Hence, the user code in the consumer function should forward this error to the corresponding entity (i.e., the orchestrator or the driver function) that can re-invoke the producer with the same original arguments.  For example, with the AWS Step Functions orchestrator, the user can define a custom fallback function to handle particular errors~\cite{aws:step-errors}.

As an alternative to offloading \shortn{} timeout error handling, providers may consider modifying their runtime (which is aware of the grace period before the instance shutdown) to back up the remaining non-retrieved objects to cloud storage transparently to the application code, by converting \shortn{} references into the provider's storage service keys. For example, when the cluster manager asks the producer instance to shut down (e.g., Knative instances have at least one minute to shut down~\cite{redhat:knative-scaling}), the instance can store remaining, non-consumed \shortn{} objects in the storage service before gracefully terminating. Unaware of the producer instance shutdown, the consumer instance can first attempt a regular \shortn{} retrieval followed by a retrieval from the storage service using the same key if the first retrieval returns an error.  If an application repeatedly faces \shortn{} errors, the infrastructure can disable \shortn{} for these functions.


\vspace{0.1in}
To summarize, \shortn{} is fully compliant with the existing at-most-once semantics of serverless function invocations and can be enhanced to at-least-once semantics using existing serverless infrastructure by introducing error handling in application logic. Alternatively, these errors can be handled by the runtime with minimal modifications. 

\section{Implementation}
\label{sec:impl}

We prototype \shortn{} in vHive~\cite{ustiugov:benchmarking}, an open-source framework for serverless experimentation that is representative of production clouds. vHive features the Knative programming model~\cite{knative} where a function is deployed as 
a containerized HTTP server's handler
(further referred to as {\em function server}), which is triggered upon receiving an HTTP request, i.e., RPC, sent to a URL assigned to the function by Knative. Each function instance runs in a separate Kubernetes pod atop a worker host (bare-metal or virtualized) in a serverless cluster.

\begin{figure}[t!]
    \centering
    \includegraphics[width=\columnwidth]{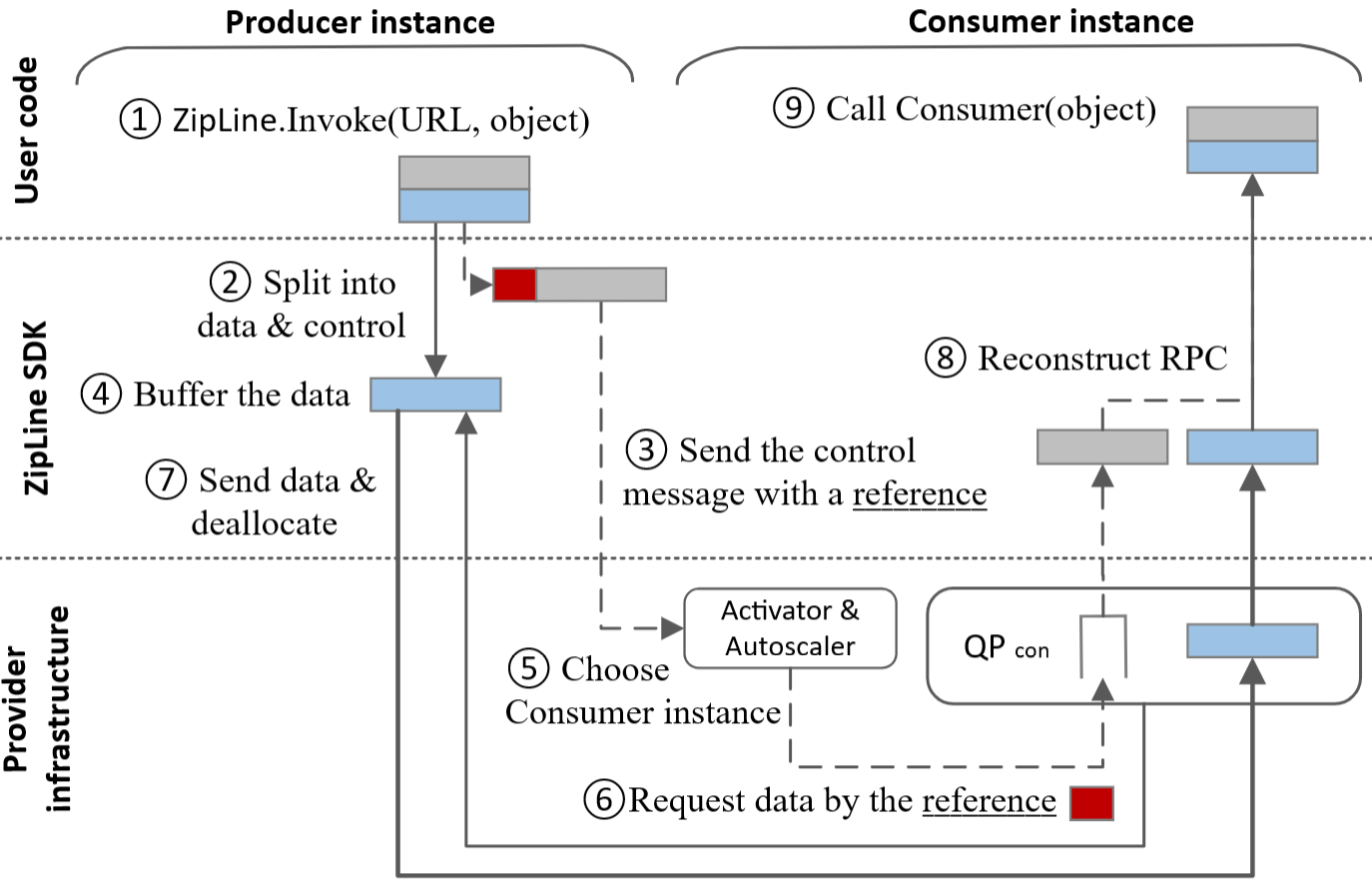}
    \caption{\shortn{} operation in a single producer single consumer scenario (only the request path is shown). Dashed arrows show the control plane, solid lines show the data plane, and the thick solid lines show data streaming in the data plane.
    }
    \label{fig:xdt-impl}
    \vspace{-15pt}
\end{figure}

\subsection{\shortn{} Prototype in vHive/Knative}
\label{sec:design-prototype}

We start by describing the implementation of the different software layers of the prototype, required to support blocking function invocations with \shortn, followed by a discussion of support for the non-blocking \shortn{} API.


\subsubsection{\shortn{} Software Development Kit (SDK)}
\label{sec:impl-sdk}

\shortn{} relies on an SDK to implement the API, bridging the user logic and the provider components that perform the transfer. At the producer instance's side, the SDK splits the original invocation request into two messages, namely a control message and an object, which comprises the transferred data. The SDK creates and adds a \shortn{} reference to the gRPC request as an HTTP header. The reference comprises an encrypted string, containing the IP address of the pod where instance's function server is running, and the object key, which is unique for that pod. Encryption prevents the user code from obtaining the IP addresses of function instances.

At the consumer instance's side, the SDK reconstructs the original request, joining the control message and the object (after the latter has been pulled), before invoking the consumer function in the same way as with the vanilla serverless API. 

\subsubsection{Control and Data Planes}
\label{sec:impl-planes}


\shortn{} uses gRPC~\cite{google:grpc} for the control plane, a common industry choice that retains compatibility to the rest of HTTP-based control-plane components. For the data plane, we choose the high-performance Cap'n Proto~\cite{capnproto} RPC fabric. This fabric runs directly on top of TCP, delivering higher performance when compared to gRPC, whose performance is limited by HTTP compatibility. Both protocols support a wide range of programming languages.

\subsubsection{Provider Components Extension}

We extend the Knative queue proxy~(QP) for object buffering~(\S\ref{sec:serverless_101}).
QP is an auxiliary provider container.
It is deployed per function instance and shares the pod with the function server. 
The added logic increases the QP memory footprint by 2MB.


Because a QP, being a minimal provider container, might be online long before the function server during a cold start, we deploy the following performance optimization. We let the QP retrieve the object on behalf of the consumer function server, instead of the consumer SDK, to overlap retrieving the request with booting the function instance.


\subsection{\shortn{} Operation}
\label{sec:xdt-operation}

\subsubsection{\shortn{} \texttt{invoke()} Operation}

Figure~\ref{fig:xdt-impl} shows the request path in the \shortn{} infrastructure following an \texttt{invoke()} call.
\circled{1} when the caller function needs to call another function it invokes the SDK.
\circled{2} the SDK splits the request into two parts, the \shortn{} object and the control plane message that carries the reference to the object.
\circled{3} the SDK sends the control message to the activator and \circled{4} stores the object into a buffer to be fetched later by the consumer's QP~(\rpcon{}).
\circled{5} the activator chooses the instance of the consumer and forwards the control message to the consumer's QP~(\rpcon{}).
\circled{6} 
\rpcon{} extracts the reference from the header, decrypts the reference to extract IP address and the object key, and requests the data by sending a Cap’n Proto RPC request to the producer function's SDK, requesting the data by the object key.
\circled{7} SDK at the producer function sends the data to the \rpcon{} and de-allocates the object when they are dispatched.
\circled{8} \rpcon{} forwards the object to the SDK that reconstructs the original request, and \circled{9} invokes the function handler.
If the response is small, it follows the reverse control plane path through the two QPs and the activator.


\subsubsection{\shortn{} \texttt{get()} / \texttt{put()} Operation}

Whereas \texttt{invoke()} is a synchronous call, the two other calls of the \shortn{} API -- \texttt{put()} and \texttt{get()} -- are asynchronous. While the operation of \texttt{put()} and \texttt{get()} is similar to \texttt{invoke()}, there are a few important differences. 
The first difference is that \texttt{put()} returns a \shortn{} reference for the object to the user logic.
The producer function may pass this reference, like any other string field, to any function that belongs to the same user. 
Once the consumer function calls \texttt{get()} using the delegated reference, the SDK retrieves the object by sending a Cap’n Proto RPC request directly to the producer instance (i.e., to a a Cap’n Proto RPC server inside the SDK), using the IP address and the key in the reference.


The asynchronous \texttt{get()/put()} API can be used not only for invocations  but also for large responses as well.
The response path follows the control plan path in the reverse order and is used only with small (inline) replies, i.e., $<$6MB in AWS Lambda. 
In the case of a large reply, the \shortn{}-enabled consumer creates a reference to the response object through a \texttt{put()} call and includes the reference in the response.
Upon receiving the response, the producer can retrieve the response payload through a \texttt{get()} call.

\subsection{Flow Control}
The \shortn{} design relies on the availability of the pre-allocated buffer in the \rpcon{} component to offer high-performance data transfers. If buffers are unavailable, the system needs to engage a flow control mechanism to pace the sending components before the downstream buffers free up. Fortunately, a Cap’n Proto RPC works on top of TCP and can rely on its flow control without any changes to the \shortn{} logic, which only needs to buffer and forward the object's chunks along the component chain. Hence, if the number of transmitted objects exceeds the number of available buffers, the subsequent transfers are paused, resulting in the user code blocking in the corresponding \shortn{} API call.

\section{Methodology}
\label{sec:method}

\subsubsection{Evaluation Platform}

We prototype and evaluate \shortn{} in Knative~\cite{knative}, which is widely used in commercial offerings~\cite{knative-offerings} and also representative of the leading close-source serverless platforms~\cite{liu:gap}. We deploy a Knative cluster that features \shortn-enabled queue-proxy containers on AWS EC2 nodes, similarly to prior work~\cite{klimovic:pocket,mahgoub:sonic,wawrzoniak:boxer}, thus ensuring low access time to AWS S3.
We use a multi-node cluster of \texttt{m5.16xlarge} instances in the `us-west-1` availability zone, to evaluate the baseline and the \shortn{}-enabled settings. This instance features Intel Xeon Platinum 8000 series 3.1GHz with 64 SMT cores, 256GB RAM, EBS storage, and a 20Gb/s NIC.

Using the vHive experimentation framework v1.4.2~\cite{ustiugov:benchmarking}, we set up Knative 1.3 in a multi-node Kubernetes 1.23 cluster~\cite{k8s}, running all deployed functions, Knative autoscaling components, and Istio ingress~\cite{istio}. The pods are scheduled on nodes to ensure all the data transfers happen across the network (i.e., no local communication), placing each function on a separate AWS EC2 node. 
In all experiments, we emulate a stable serverless workflow where active instances are always present -- i.e., there are no cold starts during the measurements. 

\subsubsection{Measurement Framework}

We use the measurement infrastructure integrated with the vHive framework~\cite{vhive:methodology}, which supports end-to-end benchmarking.
The vHive framework features a service, called invoker, that injects requests in a common format for all of the studied workloads and waits for the responses from the corresponding workflows, reporting the end-to-end delays.
The user code of workloads is annotated with logs, which are then aggregated to determine the end-to-end latency breakdown. 
Unless specified otherwise, we report average end-to-end latency based on 10 measurements. For microbenchmarks, which do not have any computational overheads except network processing, we calculate {\em effective bandwidth} of a data transfer by dividing the transferred object size by the measured end-to-end latency.

\subsubsection{Baseline and \shortn{} Configurations} 
Our baseline is the through-storage communication approach.
We evaluate two options for the storage service that represent the lower and higher performance {\em and} cost bounds of the previously proposed multi-tier ephemeral storage solutions~\cite{klimovic:pocket,sreekanti:cloudburst,mahgoub:sonic,romero:faast,pu:shuffling}. The first is Amazon S3, which is the baseline configuration used in prior work and represents the cheapest, albeit slowest, storage option in today's clouds.
The second is ElastiCache, a cloud-native in-memory data store that offers the fastest, albeit most expensive, storage solution today, >100\tms{} expensive compared to S3. \footnote{We configure ElastiCache in the on-demand mode, which is $>$4\tms{} cheaper than its newly-added serverless mode~\cite{elasticache:pricing}.}
As shown in prior work~\cite{klimovic:understanding,klimovic:pocket}, ElastiCache provides extremely high performance for inter-function communication but at a high monetary cost as compared to S3.
For ElastiCache, we used a single-node Redis cache of the node type \texttt{cache.m6g.16xlarge} having 64 vCPUs with 25 Gb/s NIC, which is one of the peak performance configurations priced at \$4.7 per hour.

\subsubsection{Microbenchmarks} 
\label{sec:method_ubenches}
We use a number of microbenchmarks, implemented in Golang 1.18, each of which evaluates one of the data transfer patterns commonly used in serverless computing~(\S\ref{sec:xdt-api}), namely producer-consumer~(1-1), scatter, gather, and broadcast. All these patterns comprise various numbers of instances of the producer and the consumer functions communicating one or more objects from the former to the latter. 
From here on, by saying a producer (consumer), we mean a producer (consumer) function instance.

\subsubsection{Real-World Workloads}
\label{sec:method_apps}

We use three data-intensive applications from the vSwarm benchmarking suite~\cite{vswarm}, which features representative workloads widely used in serverless computing, with their reference inputs. Each workload is comprised of multiple functions, deployed with Knative Serving~\cite{knative:serving}, that call one another using the blocking interface, i.e., a caller function waits for the callee to respond. Each of the workloads uses one or more data transfer patterns to communicate across functions. We modify the workloads to support \shortn{} along with the S3-based and ElastiCache baselines using the same communication API: \texttt{invoke()}, \texttt{get()} and \texttt{put()} (\S\ref{sec:xdt-api}).

The studied workloads have different communication patterns. {\em Video Analytics (VID)} shows the 1-1 and scatter patterns, as it features a pipeline of video streaming, frame decoder, and object recognition functions; where the frame decoder function invokes the object recognition function once for several frames in a decoded fragment in the scatter communication pattern. {\em Stacking Ensemble Training (SET)} is a distributed ML training application that fits the serverless programming model well due to its speed, low memory footprint, and low computational complexity~\cite{stacking1,stacking2,stacking3}.
In this workload, the first function broadcasts the training dataset when invoking several training tasks in parallel, and the last function gathers and reconciles the trained ensemble model. Hence, this workload's execution highly depends on the efficiency of the broadcast and gather communication patterns. Finally, the {\em MapReduce (MR)} workload implements the Aggregation Query from the representative AMPLab Big Data Benchmark~\cite{pavlo:comparison}. The gather pattern's execution is critical for the MapReduce workload, due to the data-intensive shuffling phase between the mapper and the reducer functions.

\subsubsection{Cost Model}
\label{sec:cost_model}

We estimate the cost of executing the applications we study from the application developer's perspective, according to the AWS pricing models~\cite{aws:pricing,s3:pricing,elasticache:pricing}. The price of a single function invocation, from the perspective of an application developer, comprises a small fixed fee for invoking a function, another fee proportional to the product of the \texttt{processing time} and \texttt{maximum memory footprint} of that invocation, and the cost of storage for transferring the data. For all studied functions, we assume the maximum function memory footprint of 512MB, and use the processing times as measured in \S\ref{sec:workload_perf}. Storage costs are billed on the \texttt{GB/month} (AWS S3~\cite{s3:pricing}) or \texttt{GB/hour} (on-demand AWS ElastiCache~\cite{elasticache:pricing}) bases. In our cost model, we take the minimal possible price for storing transferred data, assuming that ephemeral storage de-allocates transferred data immediately after the last retrieval. Note, this is rarely the case in today's production systems: e.g., AWS S3 lacks support for expiration time below one day as of 2024, effectively leaving expired objects cleaning to the application.

\section{Evaluation}
\label{sec:eval}

We compare \shortn{} to through-storage transfers based on Amazon S3 and ElastiCache (EC). We first study the performance of the evaluated communication mechanisms on microbenchmarks. We then assess the performance and cost of real-world applications running in a serverless cloud. 


\subsection{Microbenchmarks}
\label{sec:ubenches}

Next, we quantify the latency and effective bandwidth of \shortn{} in common communication scenarios~(\S\ref{sec:method_ubenches}): 1-1, gather, scatter, and broadcast.
The 1-1, or producer-consumer, pattern is typical of chained function invocations accomplished via the \texttt{invoke()} API call. 
Gather, or reduce, is essential for applications with functions whose input is the output of several other functions and that use the \texttt{put()}/\texttt{get()} API.
Scatter, or map, is important when functions have a large fan-out of calls to other functions, passing the objects via the \texttt{invoke()} and the \texttt{put()}/\texttt{get()} APIs. 
Broadcast is used by functions that distribute the same data among many consumers, accomplished via a single \texttt{put()} call followed by multiple \texttt{get()} calls with the {\em same} S3 key or \shortn{} reference.

\subsubsection{Producer-Consumer Communication}

\begin{figure}[t]
\subfloat[Latency CDFs for 10KB objects.] {
        \includegraphics[width=0.49\columnwidth]{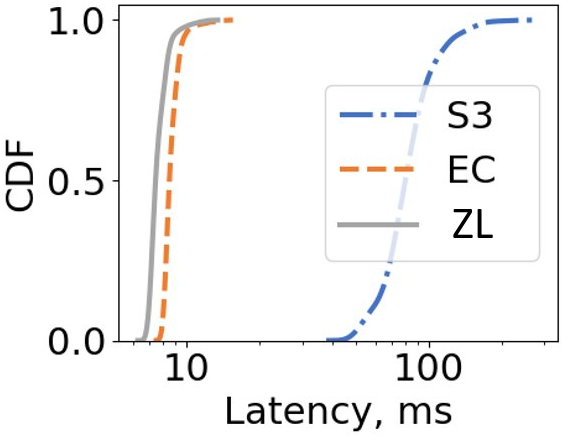}
        \label{fig:tail_cdfs}}
\centering
\subfloat[Latency CDFs for 10MB objects.] {
        \includegraphics[width=0.49\columnwidth]{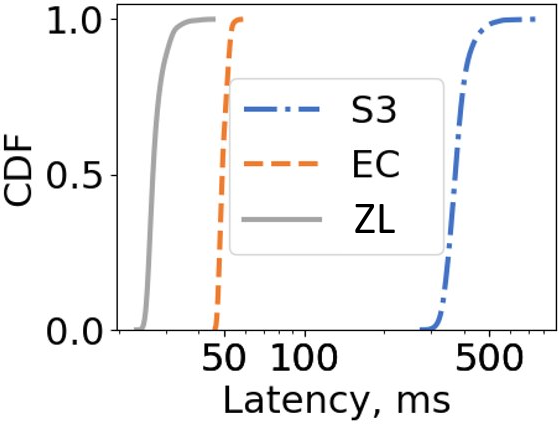}
        \label{fig:tail_sweep}}
\newline
\centering
\caption{Transfer latency cumulative distribution functions~(CDFs) for S3, ElastiCache (EC) and \shortn{} in the {\em 1-1} workflow. Note the log. scale on the horizontal axis.
}
\vspace{-15pt}
\label{fig:eval_tail}
\end{figure}

We focus on the 1-1 (producer-consumer) pattern to study the latency characteristics of the communication methods. Latency is a key metric for interactive, user-facing cloud services, with both median and tail latency considered critical. 

Figure~\ref{fig:eval_tail} plots the median and tail (99th percentile) latency for S3, ElastiCache and \shortn{}-based transfers for 10KB (small) and 10MB (large) objects. For small objects, transfers through ElastiCache in-memory cache offer much lower latency than transfers through S3, a cold storage service. The median (tail) latency with ElastiCache is 89\% (92\%) lower than that with S3. \shortn{} offers a further improvement compared to ElastiCache, with median (tail) latency 12\% (10\%) lower than ElastiCache. \shortn{} has better latency than transfers through S3 and ElastiCache because \shortn{} avoids writing and reading the object on intermediate nodes. 
For large objects, the median (tail) latency of the ElastiCache-based transfers is 87\% (90\%) lower than the S3-based ones.
\shortn{} shows median and tail transfer latencies 45\% and 34\% shorter than those with ElastiCache. Larger object sizes incur higher write and read latencies while transferring the objects through third-party services, which explains the performance advantages of \shortn{} over both S3 and ElastiCache. 





\begin{figure}[t]
\subfloat[10KB object transfers] {
        \includegraphics[width=\columnwidth]{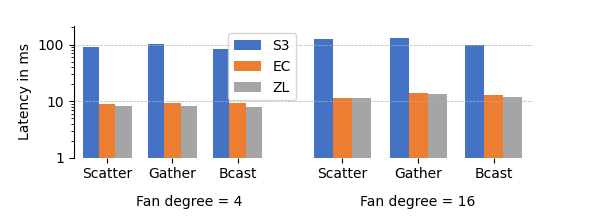}
        \label{fig:eval_BW-10KB}}
\newline
\centering
\subfloat[10MB object transfers] {
        \includegraphics[width=\columnwidth]{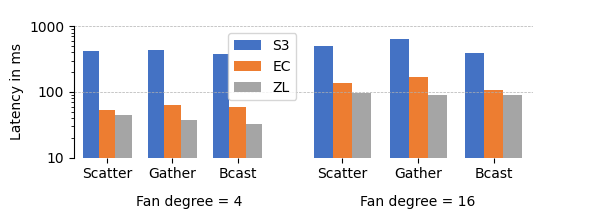}
        \label{fig:eval_BW-10MB}}
\newline
\centering
\caption{Transfer latency of the scatter, gather, and broadcast communication patterns with the fan degrees of 4 and 16. Note that both subfigures use a logarithmic scale on the vertical axis, but the scales differ across subfigures.}
\label{fig:eval_BW}
\end{figure}

\subsubsection{Collective Communication}
\label{sec:eval_bw}

We evaluate the speed of the collective communication patterns, namely the gather, scatter, and broadcast, by comparing their latency and effective bandwidth, calculated as the size of the transferred objects divided by the end-to-end transfer time.
We study fan-in (gather) and fan-out (scatter, broadcast) degrees of 4 and 16, and consider 10KB (small) and 10MB (large) transfer sizes.

Figure~\ref{fig:eval_BW-10KB} shows the results for S3, ElastiCache, and \shortn{} transfers of 10KB. For the small transfers, ElastiCache consistently outperforms S3, delivering a latency 9.2-11.0$\times$ lower at the fan degree of 4 and 7.8-10.8$\times$ lower at the fan degree of 16. This result corroborates prior work~\cite{klimovic:pocket} that also noted that transfers via in-memory storage such as ElastiCache significantly improve performance over transfers via S3. \shortn{} consistently matches or outperforms ElastiCache, with a latency up to 1.16$\times$ lower than ElastiCache. 

These trends persist for larger 10MB transfers as well, shown in Figure~\ref{fig:eval_BW-10MB}). ElastiCache continues to outperform transfers through S3, with the transfer latency up to 7.7$\times$ lower. Meanwhile, \shortn{} improves on ElastiCache by delivering 1.2-1.9$\times$ lower latency. 

Lastly, \shortn{} achieves higher effective bandwidth than S3 and ElastiCache. For 10MB transfers with a fan degree of 32, \shortn{} achieves 16.4Gb/s (82\% of the NIC peak bandwidth of 20Gb/s). In contrast, ElastiCache-based transfers deliver 14.0Gb/s (70\% of the peak bandwidth) while S3-based transfers deliver 5.5Gb/s (28\% of the peak bandwidth).  

\subsection{Real-World Workloads}

\label{sec:workload_perf}

\begin{figure}[t]
\subfloat[Video Analytics] {
        \includegraphics[width=0.89\columnwidth]{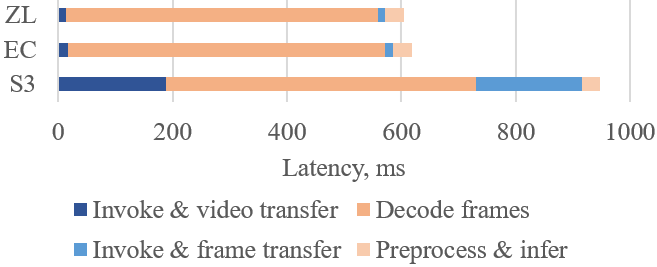}
        \label{fig:video}}
\newline
\centering
\subfloat[Stacking Ensemble Training] {
        \includegraphics[width=0.89\columnwidth]{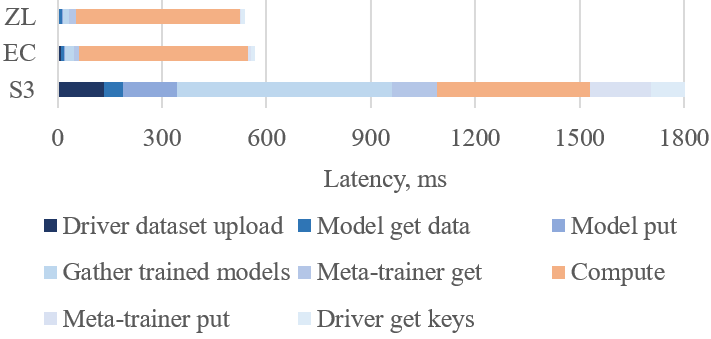}
        \label{fig:stacking}}
\newline
\centering
\subfloat[MapReduce] {
        \includegraphics[width=0.89\columnwidth]{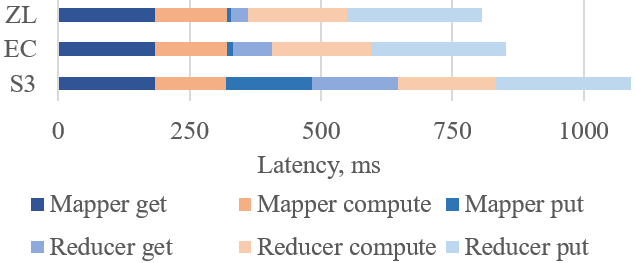}
        \label{fig:stacking}}
\newline
\centering
\caption{Latency breakdown of real-world workloads, deployed in \shortn{}, ElastiCache (EC) and S3 based systems.}
\vspace{-15pt}
\label{fig:workloads}
\end{figure}

Next, we study three data-intensive applications (\S\ref{sec:method_apps}), presenting their end-to-end latency along with a detailed breakdown of the sources of latency (Figure~\ref{fig:workloads}) and estimating the associated cost (Table~\ref{tbl:cost}) of executing an invocation for each of the studied applications.


\begin{table}
\centering
\arrayrulecolor{black}
\resizebox{\columnwidth}{!}{%
\begin{tabular}{!{\color{black}\vrule}l!{\color{black}\vrule}r!{\color{black}\vrule}r!{\color{black}\vrule}r!{\color{black}\vrule}r!{\color{black}\vrule}r!{\color{black}\vrule}r!{\color{black}\vrule}r!{\color{black}\vrule}} 
\cline{2-8}
\multicolumn{1}{l!{\color{black}\vrule}}{} & \multicolumn{3}{c!{\color{black}\vrule}}{S3}                                                                                                        & \multicolumn{3}{c!{\color{black}\vrule}}{ElastiCache}                                                                                                        & \multicolumn{1}{l!{\color{black}\vrule}}{\textbf{\shortn{}}}            \\ 
\hline
App                                        & \multicolumn{1}{l!{\color{black}\vrule}}{Comp.} & \multicolumn{1}{l!{\color{black}\vrule}}{Stor.} & \multicolumn{1}{l!{\color{black}\vrule}}{\textbf{Total}} & \multicolumn{1}{l!{\color{black}\vrule}}{Comp.} & \multicolumn{1}{l!{\color{black}\vrule}}{Stor.} & \multicolumn{1}{l!{\color{black}\vrule}}{\textbf{Total}} & \multicolumn{1}{l!{\color{black}\vrule}}{\textbf{Total (comp.)}}  \\ 
\hline
VID                                        & 37                                              & 18                                              & \textbf{55}                                              & 14                                              & 913                                             & \textbf{928}                                             & \textbf{17}                                                       \\ 
\hline
SET                                        & 95                                              & 30                                              & \textbf{125}                                             & 69                                              & 1104                                            & \textbf{1172}                                            & \textbf{70}                                                       \\ 
\hline
MR                                         & 180                                             & 416                                             & \textbf{595}                                             & 125                                             & 99667                                           & \textbf{99792}                                           & \textbf{129}                                                      \\
\hline
\end{tabular}}
\caption{
	Cost estimation (in $USD\times10^{-6}$) for compute (Comp) and storage (Stor) spending when executing a single invocation for S3, ElastiCache, and \shortn{} based configurations based on AWS Lambda~\cite{aws:pricing}, AWS S3~\cite{s3:pricing}, and AWS ElastiCache~\cite{elasticache:pricing} prices as of 1/1/2023.
	}
	\vspace{-15pt}
	\label{tbl:cost}
\arrayrulecolor{black}
\end{table}

\subsubsection{Video Analytics (VID)} 
\noindent\textbf{Performance.} The workload spends 39\% and 5\% of its execution time transferring the video fragment and the frames in the S3-based and ElastiCache-based configurations, respectively. With \shortn{}, this fraction decreases to 4\%, reducing the overall processing time by 36\% and 2\% vs. the S3 and ElastiCache baselines, respectively. This speedup comes from 9.5\tms{} and 1.2\tms{} faster transmission of video and frames, respectively. 

\noindent\textbf{Cost.} A single invocation processed in a \shortn{}-enabled system lowers the cost by 3$\times$ and 56$\times$, 
compared to S3 and ElastiCache-based configurations, respectively.


\subsubsection{Stacking Ensemble Training (SET)}
\noindent\textbf{Performance.} SET spends 76\% and 14\% of execution time in communication in the S3-based and ElastiCache-based configuration, respectively. The largest fraction of data communication is the {\em gather trained models} latency component, accounting for 34\% and 4\% of the overall execution time in the S3-based and ElastiCache-based configurations, respectively. Using \shortn{} decreases the gather fraction to 3\% of the end-to-end latency, driving the communication fraction down to 12\%. 
Thus, \shortn{} delivers a 3.4\tms{} speedup over the S3 baseline and 1.05\tms{} vs. ElastiCache. 

\noindent\textbf{Cost.} \shortn{} is cheaper by 2$\times$ and 17$\times$,
when compared to the S3 and ElastiCache based alternatives, respectively.


\subsubsection{MapReduce (MR)} 

\noindent\textbf{Performance.}
The workload shows 70\% and 62\% of execution time spent in communication for the S3 and ElastiCache configurations respectively. Moreover, 40\% of the overall time in S3 baseline is spent retrieving the original input from S3 and writing back the results to S3, which we do not optimize with \shortn{}. The rest, i.e., 30\% of time, are subject to \shortn{} optimization. 
\shortn{} delivers 1.26\tms{} overall speedup over the S3 baseline and 1.05\tms{}  over ElastiCache. 
\shortn{}'s speedup is due to a significant decrease in data shuffling, namely mapper-put and the reducer-get phases, which are reduced by 23.4\tms{} and 4.8\tms{}, respectively, compared to the S3 baseline, and by 30\% and 55\%, respectively, compared to ElastiCache.

\noindent\textbf{Cost.} Compared to the two previously-discussed workloads, \shortn{} reduces the cost of executing MapReduce by 5$\times$ and 772$\times$
vs. S3- and ElastiCache-based alternatives, respectively. 
This large cost reduction associated with \shortn{} is attributable to the large amount of ephemeral data transferred during the shuffle phase of MapReduce, making through-storage/cache transfers particularly expensive.


\subsection{Summary}

\shortn{} enables efficient transfer of ephemeral data across functions without adding cost or complexity to the application logic, delivering high performance, compatibility with existing semantics and API, and native autoscaling.
By design, \shortn{} is much faster than transferring data via conventional storage services, such as AWS S3. \shortn{} avoids unnecessary writing and reading to the durable tier of storage services, which incurs high latency overheads and carries a monetary cost. Compared to an in-memory cache, \shortn{} offers similar or better performance {\em without} the staggering cost or complexity overheads associated with using an additional service. Indeed, our results show that the cost overheads of an in-memory cache are prohibitive, exceeding compute costs by one to two orders of magnitude (Table~\ref{tbl:cost}). Meanwhile, using an additional service for the caching tier burdens the developer with additional design complexity in the application logic and may require manual reconfiguration (or further application complexity) to accommodate changes in load or data volume. In contrast, \shortn{} avoids the need for an additional service and its bandwidth naturally scales with the number of producer and consumer instances. 

\section{Related Work}
\label{sec:related}

Prior works~\cite{pu:shuffling,sreekanti:cloudburst,romero:faast,klimovic:pocket,mahgoub:sonic} consider several ephemeral storage service designs, aiming to provide high-performance transfers at a reasonable cost. However, we show in \S\ref{sec:storage-and-cost} that the cost of even the slowest tier (e.g., AWS S3 as in several works~\cite{klimovic:pocket,pu:shuffling,mahgoub:sonic}) can dominate the overall cost of executing a data-intensive application in serverless clouds.
Other prior works~\cite{shillaker:faasm,wang:reference,mvondo:ofc,romero:faast, khandelwal:jiffy,du:molecule,spright:sigcomm22} consider extending serverless with a distributed shared memory (DSM) tier and pass references over the DSM around instead of data objects. In contrast to these proposals, the data objects transmitted via \shortn{} are immutable, avoiding the complexity of supporting data consistency models. YuanRong~\cite{yuanrong:sigcomm24} is a Huawei production system that can pass data by references, but the paper lacks implementation details.

Other works explore support for {\em connection-based} direct communication for serverless applications~\cite{wawrzoniak:boxer, wawrzoniak:off-the-shelf, thomas:particle,nesterov:floki}, which can help to port microservice and monolith applications to serverless but is generally not used in serverless-native applications~\cite{romero:faast,eismann:state}. 
We argue that such optimizations undermine the core principle of serverless, namely the cloud provider's transparent management of cloud infrastructure, particularly autoscaling. With connection-based approaches, scaling the number of instances of a function up or down is not transparent to its producers and consumers, requiring a reconfiguration of the workflow topology to accommodate.
In contrast, \shortn{} is fully compatible with the object-centric \texttt{get()/put()} API widely adopted by applications and seamlessly works with the existing cloud autoscaling infrastructure.

Like in the \shortn{} design, researchers have proposed separating the control and data planes to avoid centralized bottlenecks and deliver high performance. 
For example, Crab~\cite{kogias:crab} and Prism~\cite{yutaro:prism} follow a similar separation to reduce the load on L4 and L7 load balancers, respectively. Dataflower~\cite{dataflower:asplos23} and FUYAO~\cite{fuyao:asplos24} adopt asynchronous transfers to decouple them from the control plane.


\shortn{} ships a function invocation's data along the compute for processing, which is complementary to the approaches that ship compute to data or data to compute. 
Shredder~\cite{zhang:shreder} suggests running compute operations directly at the storage tier. 
Other works~\cite{you:ship,ankit:adaptive,yu:following} investigate the balance between moving data vs. moving compute, suggesting hybrid schemes to combine both.

\shortn{} enables high-performance transfers without making assumptions on function instances co-location and data locality, which makes it fundamentally different to the following prior works. 
SAND~\cite{akkus:sand} accelerates data communication proposing a hierarchical messaging bus to facilitate transfers between co-located function instances. 
FaaSFlow~\cite{li:faasflow}, Sledge~\cite{lyu:sledge}, and Wukong~\cite{carver:wukong} focus on leveraging locality to accelerate the execution of serverless multi-function applications.
Nightcore~\cite{jia:nightcore} suggests exchanging messages over OS pipes for co-located functions.
Despite the potential efficiency gains, today's commercial systems, e.g., AWS Lambda, tend to avoid serverless function co-location as such placements may lead to hotspots~\cite{agache:firecracker,balaji2021fireplace}, instead relying on statistical multiplexing across a wide server fleet.

\section{Conclusion}
\label{sec:concl}


The cost and performance of data-intensive serverless applications heavily depends on the efficiency of inter-function data transfers. 
The state-of-the-art data transfer methods for serverless clouds fall short of serverless applications' demands. In response, we introduce \shortn{}, a high-speed API-preserving direct function-to-function communication method that integrates seamlessly with the existing autoscaling infrastructure. \shortn{} leverages control/data separation and references to provide low latency and high bandwidth for all typical serverless communication patterns. 
A \shortn{} prototype reduces the overall cost of executing the end-to-end application invocation by 2-5\tms{} over the through-storage transfers {\em and} accelerates real-world serverless applications by 1.3-3.4\tms{}. Compared to through-cache transfers, \shortn{} slashes the cost by 17-772\tms{} while yielding speedups of 2-5\%.



\bibliographystyle{plain}
\bibliography{./bibcloud/gen-abbrev,dblp,ref}

\end{document}